\documentclass[aps,prb,twocolumn,superscriptaddress,raggedbottom,showpacs,floatfix]{revtex4-1}
\usepackage{amsmath,amsfonts,amssymb,amsthm}
\usepackage{color}
\usepackage{graphicx,latexsym}

\begin{document}
\renewcommand{\vec}{\mathbf}
\renewcommand{\Re}{\mathop{\mathrm{Re}}\nolimits}
\renewcommand{\Im}{\mathop{\mathrm{Im}}\nolimits}

\title{Resonant manifestations of chiral excitons in Faraday and Kerr effects in topological insulator film}
\author{D.K. Efimkin}
\affiliation{Institute of Spectroscopy  RAS, 142190, Troitsk, Moscow Region, Russia}
\author{Yu.E. Lozovik}
\altaffiliation{email: lozovik@isan.troitsk.ru}

\affiliation{Institute of Spectroscopy RAS, 142190, Troitsk, Moscow Region, Russia}
\affiliation{Moscow Institute of Physics and Technology, 141700,
Moscow, Russia}

\begin{abstract}
Manifestations of chiral excitons on a magnetically gapped surfaces of a topological insulator thin film in  Kerr and Faraday effects are analyzed. Excitonic contribution to a surface optical conductivity tensor is calculated. Chiral excitons contrary to conventional ones resonantly contribute to Hall conductivity due to the lack of the symmetry between the states with opposite angular momentum.  They can lead to the considerable enhancement of Faraday angle and ellipticity of transmitted electromagnetic wave. Chiral excitons cause decrease of Kerr angle and prominent signatures in ellipticity of reflected electromagnetic wave. Conditions for experimental observation of described effects are discussed.
\end{abstract}
\maketitle

\section{Introduction}
Theoretical and experimental study of topological insulators (TI) that have nontrivial topology intrinsic to their band structure grows rapidly (see \cite{HasanKane,QiZhang} and references therein). TI have forbidden band in a bulk but on their surface (3D) or edge (2D) there are very unusual electronic states. Recently a ``new generation'' of three-dimensional topological insulators (the  compounds $\mathrm{Bi}_2\mathrm{Se}_3$, $\mathrm{Bi}_2 \mathrm{Te}_3$ and others), retaining topologically protected behavior at room temperatures, were investigated  experimentally \cite{Chen,Hsieh,Xia}. The surface states of these materials are protected from nonmagnetic disorder and obey Dirac equation for massless two-dimensional particles that is analogous to one for electrons in graphene, unique two-dimensional carbon based material with extraordinary electronic and mechanical properties \cite{GrapheneExp1,GrapheneExp2,GrapheneReview}.

Interesting physics arises when time reversal symmetry on the surface of TI is broken by external exchange field. Exchange field that can be created by ordered magnetic impurities introduced to a TI bulk \cite{TIExchange1,TIExchange2} acts only on magnetic moment of the electrons and generates the energetic gap in the surface spectrum. Contrary to the initial gapless state the set of excitonic states appears in the gap due to Coulomb interaction on the surface. It is interesting that the excitonic state has minimal energy at finite value of orbital angular momentum quantum number and can be called ``chiral''\cite{GarateFranz}.

The time reversal symmetry breaking leads to half-integer quantization of the surface Hall conductivity and, as result, to quantized  Faraday and Kerr effects on the surface of TI\cite{MaciejkoQiDrewZhang,LanWanZhang,TseMacDonaldMF,TseMacDonaldEF}.  If the time reversal symmetry is broken on the whole surface of TI low frequency electromagnetic response of TI bulk can be described by \emph{macroscopic} approach based on Lagrangian for electromagnetic field with additional  $\theta$-term that corresponds to topological magneto-electric effect\cite{QiHughesZhang,EssinMooreVanderbilt}. The topological magneto-electric effect in TI bulk provide a solid state realization of axion electrodynamics \cite{Wilczek}. In thin film of TI  which width is considerably smaller then length of incident electromagnetic wave Faraday angle $\theta_\mathrm{F}$ and Kerr angle $\theta_\mathrm{K}$ are universal: $\tan \theta_\mathrm{F}=\alpha$ and $\tan \theta_\mathrm{K}=1/\alpha$, where $\alpha\approx1/137$ is fine structure constant.  Within {\it macroscopic} approach the roles of oblique incidence of electromagnetic wave\cite{LanWanZhang},  substrate\cite{MaciejkoQiDrewZhang} and interference in thick TI film \cite{LanWanZhang}  were theoretically investigated.  In spite of its mathematical elegancy this approach is well justified only in low frequency limit and does not allow to take into account the frequency dispersion effects and many-body correlations on the TI surface.

There is another approach for investigation of the  effects on TI surface based on the Maxwell equations. Response of the TI surface in these approach is characterized by the {\it microscopically} calculated its optical conductivity tensor.  The role of frequency dispersion of the optical conductivity was investigated for TI film subjected to external exchange field \cite{TseMacDonaldEF}. Also the inverse Faraday effect that manifests itself as generation of spin polarization under illumination of circularly polarized electromagnetic wave was considered \cite{MisawaYokoyamaMurakami}. But the manifestation of collective excitations --- excitons in Faraday and Kerr effects has not been considered before.

Here we theoretically investigate the role of chiral excitons on a TI surface with magnetically induced gap in  Faraday and Kerr effects.

The rest of the paper is organized as follows.  In Section 2 we briefly discuss the electronic structure of a TI surface with magnetically induced gap. Section 3 is devoted to descriptions of chiral excitons. In Section 4 the contribution of chiral excitons to optical conductivity is calculated. Section 5 is devoted to Faraday and Kerr effects in thin film of TI. Section 6 is devoted to conclusions.

\section{Electronic structure of TI}
Electrons populating the surface states of TI in presence of the external exchange field can be described by the following single particle Hamiltonian\cite{HasanKane,QiZhang}
\begin{equation}
\label{HamSP}
H_0=v_\mathrm{F}\vec{n}[\vec{k}\times \mathbf{\sigma}]+\Delta\sigma_z,
\end{equation}
where $v_\mathrm{F}$ is the Fermi velocity of electron; the vector $\mathbf{\sigma}=(\sigma_x, \sigma_y)$ consists of Pauli matrices acting in the space of its spin projections;  $\Delta$ parameterize the coupling of $z$-component of the exchange field to of electron's spin. Other components of exchange field can be excluded by gauge transformation that shifts Dirac point in momentum space. It can be showed that the magnetic field caused by a layer of ordered magnetic impurities is small and its effect on the Dirac electrons can be neglected. The spectrum  $E_{\vec{k}\gamma}=\gamma\epsilon_\vec{k}=\gamma \sqrt{|\Delta|^2+v_F^2k^2}$ is formed by conduction ($\gamma=1$) and valence ($\gamma=-1$) bands separated by the gap $2|\Delta|$. Corresponding eigenfunctions of the Hamiltonian (\ref{HamSP}) can be written as $e^{i\vec{k}\vec{r}} |f_{\vec{k}\gamma}\rangle$, where $|f_{\vec{k}\gamma}\rangle$ is a spinor part of the wave function:
\begin{equation}
|f_{\vec{k}\gamma}\rangle=\left(\begin{array}{c}
\cos(\theta_{\vec{k}\gamma}/2) e^{-i\varphi_\vec{k}/2}\\
i \sin(\theta_{\vec{k}\gamma}/2) \gamma e^{i\varphi_\vec{k}/2}\end{array}\right),
\end{equation}
where $\cos(\theta_{\vec{k}\gamma})=\gamma \Delta/\epsilon_{\vec{k}}$, and $\varphi_\vec{k}$ is polar angle for momentum vector $\vec{k}$.

A starting point for description of chiral excitons is the many-body Hamiltonian describing interacting electrons on the surface of TI
\begin{equation}
\label{ManyBodyHam}
\begin{split}
&H=\sum_{\mathbf{k}\gamma}\epsilon_{k\gamma}a_{\mathbf{k}\gamma}^+a_{\mathbf{k}\gamma} +\frac{1}{2}\sum_{\begin{smallmatrix}\mathbf{q}\mathbf{k} \\ \mathbf{k}^\prime\end{smallmatrix}}\sum_{\begin{smallmatrix}\gamma_1\gamma_2 \\ \gamma_1^\prime\gamma_2^\prime\end{smallmatrix}}\langle f_{\mathbf{k}+\mathbf{q},\gamma_1^\prime}|f_{\mathbf{k}\gamma_1}\rangle \\& \times  \langle
f_{\mathbf{k}^\prime-\mathbf{q},\gamma_2^\prime}|f_{\mathbf{k}^\prime\gamma_2}\rangle  V_\mathrm{c}(\vec{q})
 a_{\mathbf{k+q},\gamma_1^\prime}^+a^+_{\mathbf{k}^\prime-\mathbf{q},\gamma_2^\prime}
a_{\mathbf{k}^\prime\gamma_2}a_{\mathbf{k}\gamma_1},
\end{split}
\end{equation}
where $a_{\mathbf{k}\gamma}$ is the destruction operator for electron with momentum $\mathbf{k}$ from the band $\gamma$; $V_\mathrm{c}(\vec{q})=2\pi e^2/\varepsilon q$ is the two-dimensional Fourier transform of Coulomb interaction potential; $\epsilon$ is effective dielectric permittivity of the TI surface.

\section{Chiral excitons}

Coulomb interaction between the electrons populating surface states of TI can lead to formation of excitons that manifest themselves as coherent superposition of interband single-particle transitions and can be represented as bound state of an electron from the conduction band and a hole from the valence band. Creation operator of exciton $d^+_{
\vec{q}}$
with center of mass momentum
$\vec{q}$ can be written as \cite{Egri}
\begin{equation}
\label{CreatOperExciton}
d^+_
\vec{q} =\sum_{k}C_{\vec{k}\vec{q}}a_{\vec{
k}+\vec{q},1}^+ a_{\vec{k},-1},
\end{equation}
where the set of coefficients $C_{\vec{kq}}$ forms the wave function of electron and hole forming exciton in the momentum representation. We considered excitons with zero center of mass momentum $\vec{q}=0$ because only they are optically active. Hence, momentum index $\vec{q}$ will be omitted below.

Within the equation of motion based approach\cite{Egri,GarateFranz} excitons can be represented as composite bosons with corresponding commutation relation $[d,d^+]=1$ and their creation operator satisfies equation
of motion $[H, d^+] = \Omega\, d^+$, where $\Omega$ is exciton energy and $H$ is Hamiltonian of interacting electrons (\ref{ManyBodyHam}). If the part of Coulomb interaction in (\ref{ManyBodyHam}) that corresponds to scattering of electrons within single band is treated within Hartree-Fock approximation the equation of motion for the excitonic creation operator leads to
\begin{equation}
\label{ExcitonEquation}
(2\epsilon_{\vec{k}} + \Sigma_{\vec{k}}^\mathrm{eh}) C_{\vec{k}}+\sum_{\vec{k}^\prime} V_{c}(\vec{k}-\vec{k}^\prime)\Lambda_{\vec{k},\vec{k}^\prime} C_{\vec{k}^\prime}=\Omega C_{\vec{k}},
\end{equation}
where $\Sigma_{\vec{k}}^\mathrm{eh}$ is the self-energy of the electron-hole pair and $\Lambda_{\vec{k},\vec{k}^\prime}$ is angular factor that are given by
\begin{equation}
\label{ExcitonPotential}
\begin{split}
\Sigma_{\vec{k}}^\mathrm{eh}=\sum_{\vec{k}^\prime} & V_c(\vec{k}-\vec{k}^\prime) \frac{\Delta^2+v_{\mathrm{F}}^2(\vec{k}\cdot\vec{k}^\prime)}{\epsilon_{\vec{k}}\epsilon_{\vec{k}^\prime}}, \\
\Lambda_{\vec{k},\vec{k}^\prime} = \frac{1}{2} \frac{v_{\mathrm{F}}^2 k k^\prime}{\epsilon_\vec{k} \epsilon_{\vec{k}^\prime}} + & \frac{1}{2}\left(1+\frac{\Delta^2}{\epsilon_\vec{k} \epsilon_{\vec{k}^\prime}}\right)\cos (\phi_{\vec{k}}-\phi_{\vec{k}^\prime}) + \\ + \,\, &\frac{i}{2} \left(\frac{\Delta}{\epsilon_\vec{k}}+\frac{\Delta}{\epsilon_{\vec{k}^\prime}}\right)\sin(\phi_{\vec{k}}-\phi_{\vec{k}^\prime}).
\end{split}
\end{equation}
The equation (\ref{ExcitonEquation}) was derived and solved numerically in \cite{GarateFranz}. The set of excitonic states in the surface gap has unusual dependence on the orbital angular momentum $m$ and can be called "chiral". A chiral exciton has minimal energy at finite value of orbital angular momentum $m=1$ and there is no symmetry between chiral excitonic states with opposite angular momenta. The sign of the orbital angular momentum that corresponds to the lowest-energy state depends on sign of $\Delta$ and hence on direction of the exchange field.  Chiral excitons also appear\cite{ParkLouie} in bilayer graphene gapped by external electric field \cite{BilayerGapTheor,BilayerGapExp1,BilayerGapExp2}. In bilayer graphene chiral excitons with minimal energy has orbital angular momentum $m=2$.

Here we develop analytical approximate solution of (\ref{ExcitonEquation}). Self-energy $\Sigma_\vec{k}^\mathrm{eh}$ only renormalizes parameters of single particle spectrum $\Delta,v_{\mathrm{F}}$. If we denote by  $\Delta,v_{\mathrm{F}}$ the renormilized parameters of the spectrum then the self-energy term in (\ref{ExcitonEquation}) can be omitted. Angular factor $\Lambda_{\vec{k},\vec{k}^\prime}$ and single particle spectrum $\epsilon_{\vec{k}}$ contain the single scale $k_\mathrm{\mathrm{\Delta}}=|\Delta|/v_\mathrm{F}$ that corresponds to crossover between linear and parabolic regimes of massive Dirac spectrum. If localization length $k_{\mathrm{exc}}$ of wave function of the relative motion in momentum space $C_{\vec{k}}$ satisfies condition $k_{\mathrm{exc}}\ll k_{\mathrm{\Delta}}$ single particle energy and the angular factor can be approximated in following way $\epsilon_\vec{k}\approx |\Delta|+v_\mathrm{F}^2 k^2/2|\Delta|$ and $\Lambda_{\vec{k},\vec{k}^\prime}\approx\cos(\phi_{\vec{k}}-\phi_\vec{k}^\prime)+i\sin(\phi_\vec{k}-\phi_\vec{k}^\prime)$. In this limit the equation (\ref{ExcitonEquation}) coincides with Schrodinger equation for 2D hydrogen atom in which: 1) multipole momenta of Coulomb potential Fourier transform are {\it shifted} by $\delta m=1$ due to the angular factor $\Lambda_{\vec{k},\vec{k}^\prime}=e^{i(\phi_{\vec{k}}-\phi_\vec{k}^\prime)}$; 2) effective reduced mass of electron in 2D hydrogen atom problem is $\mu^\star=|\Delta|/2v_\mathrm{F}^2$. The chiral excitons can be characterized by radial $n=0,1,...$ and orbital angular $m=0,\pm1,...$ quantum numbers with $|m-1|\le n$. Their energy spectrum $\Omega_{nm}$ and wave functions in momentum space $C_k^{nm}$ can be obtained by {\it shift} of well-known ones for a 2D hydrogen atom. Energy of excitonic level $|n,m\rangle$ is given by
\begin{equation}
\Omega_{nm}=2|\Delta|-\frac{\alpha_\mathrm{c}^2 |\Delta|}{(2n+1)^2},
\end{equation}
where $\alpha_\mathrm{c} = e^2/\epsilon \hbar v_\mathrm{F} $ is dimensionless coupling strength.
State $|0,1\rangle$ with minimal energy  has orbital angular momentum $m=1$. Bohr radius of the chiral exciton and corresponding Rydberg energy are $(\alpha_\mathrm{c} k_\mathrm{\Delta})^{-1}$ and $\alpha_\mathrm{c}^2 |\Delta|$.
Characteristic momentum $k_{\mathrm{exc}}$ of excitonic wave function $C_{\vec{k}}$ localization in momentum space can be estimated as inverse Bohr radius and is equal to $k_{\mathrm{exc}}=\alpha_\mathrm{c} k_\mathrm{\Delta}$. Therefore the shifted eigenstates of the Schrodinger equation for a 2D hydrogen atom are approximate solutions of (\ref{ExcitonEquation})  if $k_{exc}\ll k_{\Delta}$ that corresponds to $\alpha_\mathrm{c}\ll1$.

\begin{figure}[t]
\label{Fig1}
\includegraphics[width=8.5cm]{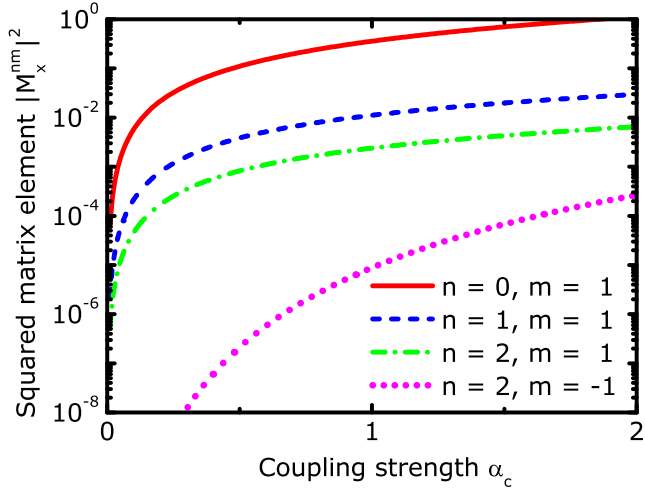}
\caption{(Color online) Squared dimensionless matrix element $|M_{x}^{nm}|^2$ for four lowest-energy optical active excitonic states ($|0,1\rangle$, $|1,1\rangle$, $|2,1\rangle$ and $|2,-1\rangle$) as a function of dimensionless coupling $\alpha_{\mathrm{c}}$.}
\end{figure}

The dimensionless parameter $\alpha_\mathrm{c}$ is the only parameter that governs physics of chiral excitons. It is approximately equal to ratio between electron-hole Coulomb interaction energy and their kinetic energies. Experimentally relevant conditions\cite{Chen,Hsieh,Xia} correspond to the case $\alpha_\mathrm{c}\ll1$ therefore we use the described analytical approximation described above for calculation of the optical conductivity tensor of a TI surface.

\begin{figure}[t]
\label{Fig2}
\includegraphics[width=8.5cm]{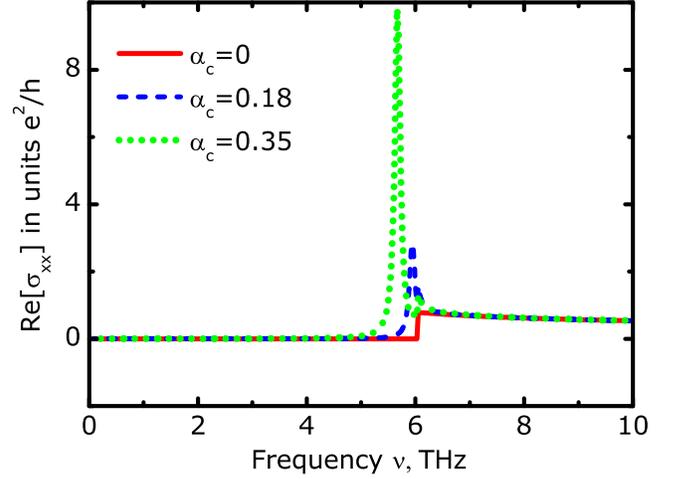}
\caption{(Color online) Frequency dependence of real part of longitudinal conductivity $\Re[\sigma_{xx}]$ for $\alpha_\mathrm{c}=0$ (solid red line), $\alpha_\mathrm{c}=0.18$ (dashed blue line) and $\alpha_\mathrm{c}=0.35$ (dotted green line).}
\end{figure}

\section{Optical conductivity tensor}
For calculation of the optical conductivity tensor of a TI surface we used the linear response theory at zero temperature. Components of the conductivity can be written in Lehmann representation in the following way
\begin{equation}
\label{ConductivityFull}
\sigma_{\alpha\beta}=\frac{e^2}{\hbar}\sum_n \frac{i}{E_{n0}}\left(\frac{J_\alpha^{n0}J_\beta^{0n}}{\omega + E_{n0}+i\delta }+\frac{J_\alpha^{0n}J_\beta^{n0}}{\omega - E_{n0}+i\delta }\right),
\end{equation}
where $|n\rangle$ and $E_{n0}$ denote excited state of the interacting system and its energy measured from the ground state $|0\rangle$ with
filled valence band and empty conduction band; $\vec{J}$ is current operator in second quantization representation. Due to momentum conservation law only states $|n\rangle$ with zero momentum contribute to optical conductivity. Corresponding two-particle states of interacting system include single-particle interband transitions $|n\rangle= a_{\vec{k}1}^+a_{\vec{k}-1}|0\rangle$ and exitonic states $|n\rangle= d^+|0\rangle$. Their contributions to the optical conductivity tensor can be separated.

After substitution of single particle sets of states $|n\rangle= a_{\vec{k}1}^+a_{\vec{k}-1}|0\rangle$ into general formula one (\ref{ConductivityFull}) can obtain the expression for the single-particle contribution to the optical conductivity tensor. The expression can be presented in Kubo-Greenwood formula form that is given by
\begin{equation}
\label{ConductivitySPKubo}
\sigma_{\alpha\beta}^{\mathrm{sp}}=\frac{e^2}{i\hbar}\sum_{\vec{k}\gamma\gamma^\prime}\frac{n_{\vec{k}\gamma}-n_{\vec{k}\gamma^\prime}}{\epsilon_{\vec{k}\gamma}-\epsilon_{\vec{k}\gamma^\prime}} \frac{\langle \vec{k}\gamma|j_\alpha|\vec{k} \gamma^\prime\rangle\langle \vec{k} \gamma^\prime|j_\beta|\vec{k} \gamma\rangle}{\omega+\epsilon_{\vec{k} \gamma}-\epsilon_{\vec{k} \gamma^\prime} + i \delta}.
\end{equation}
Here $\vec{j}=v_\mathrm{F} [\vec{\sigma}\times \vec{n}]$ is the single-particle current operator; $n_{\vec{k}\gamma}$ is occupation number of the corresponding state at zero temperature. After evaluation of (\ref{ConductivitySPKubo}) one can obtain
\begin{equation}
\label{ConductivitySP1}
\begin{split}
\Re[\sigma_{xx}^{\mathrm{sp}}]=&\frac{e^2}{h}\frac{\pi}{8}\left[1+\left(\frac{2\Delta}{\omega}\right)^2\right]\Theta(|\omega|-2|\Delta|), \\
\Im[\sigma_{xx}^{\mathrm{sp}}]=&\frac{e^2}{h}\left[\frac{|\Delta|}{2\omega}+\left(\frac{1}{8}+\frac{\Delta^2}{2\omega^2}\right)\ln\left|\frac{\omega-2|\Delta|}{\omega+2|\Delta|}\right|\right], \end{split}
\end{equation}
\begin{equation}
\label{ConductivitySP2}
\begin{split}
\Re[\sigma_{yx}^{\mathrm{sp}}]=&-\frac{e^2}{h}\times\frac{\Delta}{2\omega} \ln\left|\frac{\omega-2|\Delta|}{\omega+2|\Delta|}\right| ,\\
\Im[\sigma_{yx}^{\mathrm{sp}}]=&\frac{e^2}{h} \times \frac{\pi \Delta }{2\omega}\Theta(|\omega|-2|\Delta|).
\end{split}
\end{equation}
The obtained expressions for single-particle contribution to optical conductivity tensor are the special case of more general formulas calculated within quantum kinetic equation in \cite{TseMacDonaldEF} that takes into account disorder and finite temperature.

\begin{figure}[t]
\label{Fig3}
\includegraphics[width=8.5cm]{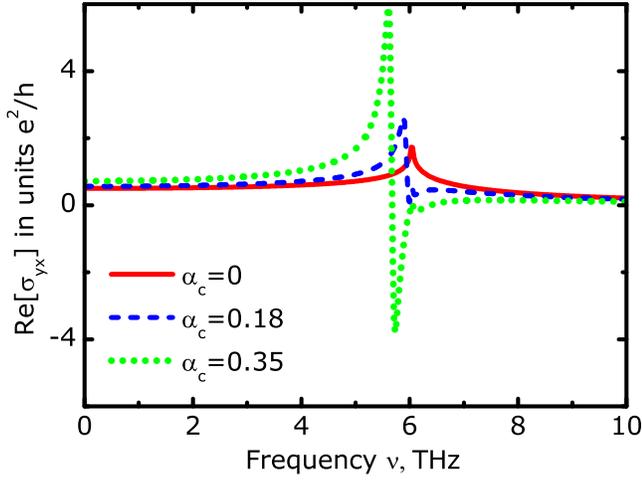}
\caption{ (Color online) Frequency dependence of real part of Hall conductivity $\Re[\sigma_{yx}]$ for $\alpha_\mathrm{c}=0$ (solid red line), $\alpha_\mathrm{c}=0.18$ (dashed blue line) and $\alpha_\mathrm{c}=0.35$ (dotted green line).}
\end{figure}

After substitution of the set of  excitonic states $d^+$ to general formula (\ref{ConductivityFull}) one can obtain the expression for the excitonic contribution to the optical conductivity tensor that have not been taken into account yet and can be presented in the following form
\begin{equation}
\sigma^{\mathrm{exc}}_{xx}=i \frac{e^2}{\hbar} \sum_{nm} |M_{x}^{nm}|^2 \frac{\omega+i\gamma}{\Omega_{nm}} \frac{2\Delta^2}{(\omega+i\gamma)^2-\Omega_{nm}^2},
\end{equation}
\begin{equation}
\sigma^{\mathrm{exc}}_{yx}=- \frac{e^2}{\hbar} \sum_{nm} m |M_{x}^{nm}|^2 \frac{2\Delta^2}{(\omega+i\gamma)^2-\Omega_{nm}^2}.
\end{equation}
 Here the summation is performed over all exciton quantum numbers; $\gamma$ is phenomenologically introduced exciton decay rate; $\vec{M}^{nm}$ is dimensionless matrix element that characterizes coupling strength of the excitonic $|n,m\rangle$ level to external electromagnetic field and depends only on dimensionless coupling strength $\alpha_\mathrm{c}$:
\begin{equation}
\label{Matrixelement}
\vec{M}^{nm}=\frac{\hbar}{\Delta} \sum_{\vec{k}}  C_{\vec{k}}^{nm} \langle k,-1|\vec{j}|k,1 \rangle.
\end{equation}
Dimensionless matrix element  $\vec{M}^{nm}$ has nonzero value only for levels with $m=\pm 1$ and all other states are optically inactive.  Dependence of squared matrix element $|M^{nm}_x|^2$ on dimensionless coupling strength for four optical active lowest-energy states is presented in Fig. 1. Absolute value of matrix element $|\vec{M}^{nm}|$ is decreasing with increasing of quantum number $n$ due to destructive interference of single-particle states.  Matrix element $\vec{M}^{nm}$ consists of coherent superposition of single particle matrix elements with weight function $C_{\vec{k}}^{nm}$ and wave functions $C^{nm}_{\vec{k}}$ for high energy excitonic levels oscillate in momentum space. Also optical activity of the excitonic levels with orbital angular momentum $m=-1$ is considerable weaker then one for the levels with $m=1$.

\begin{figure}[t]
\label{Fig4}
\includegraphics[width=8.5cm]{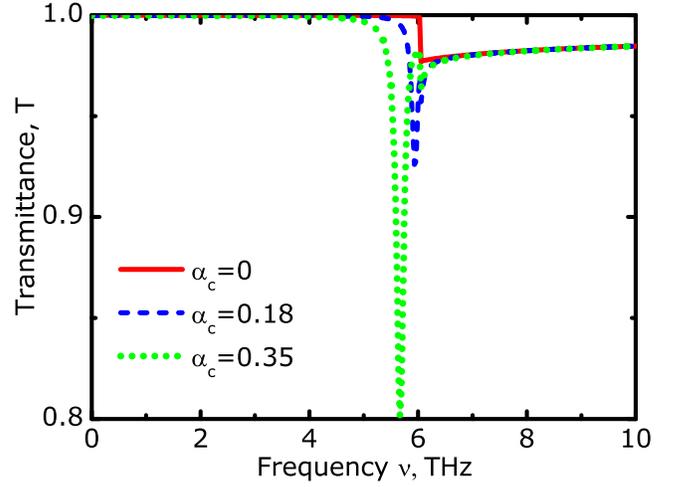}
\caption{(Color online) Frequency dependence of transmittance $T$ through TI thin film for $\alpha_\mathrm{c}=0$ (solid red line), $\alpha_\mathrm{c}=0.18$ (dashed blue line) and $\alpha_\mathrm{c}=0.35$ (dotted green line).}
\end{figure}

Contribution of the excitonic level $|n,m\rangle$ to Hall conductivity has the same sign as it orbital angular momentum quantum number $m$. States with opposite orbital angular momenta $m$ and $-m$ are connected by the time reversal transformation. In conventional two- and three- dimensional insulators there is symmetry between states with opposite angular momenta ($\Omega_{nm}=\Omega_{n-m}$ and $|M^{nm}_x|^2=|M^{n-m}_x|^2$) due to time reversal symmetry. Therefore total contribution of all excitonic states to Hall conductivity is zero.  For chiral excitons on a TI surface with magnetically opened gap the symmetry between states with opposite orbital angular momenta is broken and they {\it do} resonantly contribute to optical Hall conductivity.

For all numerical calculations we used the following parameters: $\Delta=12.5\;\hbox{meV}$, $\gamma=0.25\;\hbox{meV}$ and $v_{\mathrm{F}}=0.62 \times 10^6 \;\hbox{m/s} $. Also we used three values of dimensionless parameter $\alpha_\mathrm{c}$. Value $\alpha_\mathrm{c}=0$ corresponds to the case of noninteractiong electrons on TI surface. Values $\alpha_\mathrm{c}=0.18$ and $\alpha_c=0.35$ correspond to values of effective permittivity  $\epsilon=20.5$ and $\epsilon=10.5$, respectively. The value $\alpha_\mathrm{c}=0.18$ corresponds to $\hbox{Bi}_2 \hbox{Te}_3$ and the other values of $\alpha_\mathrm{c}$ are used for the comparison.

Real parts of longitudinal and Hall components of the optical conductivity tensor $\Re[\sigma_{xx}(\omega)]$ and $\Re[\sigma_{yx}(\omega)]$ are represented in Fig. 2 and Fig. 3, respectively. At low frequencies contribution of excitons can be neglected and conductivity tensor tends to its single-particle value (\ref{ConductivitySP1},\ref{ConductivitySP2}). Chiral excitons correspond to sharp maximum of longitudinal conductivity that leads to resonant absorption of energy of electromagnetic wave. For the used set of parameters the only maximum can be distinguished that corresponds to excitonic level $|0,1\rangle$ and contribution of other excitonic levels almost merges with contribution of single-particle transitions. Contrary to conventional ones chiral excitons resonantly contribute to Hall conductivity and, hence, can play important role in  effects.

\section{Faraday and Kerr effects in TI thin film}
We consider  effects at normal incidence of electromagnetic wave in thin film of TI which width is less then the wave length. Our results can be easily generalized to more complicated geometry and oblique incidence.

\begin{figure}[t]
\label{Fig5}
\includegraphics[width=8.5cm]{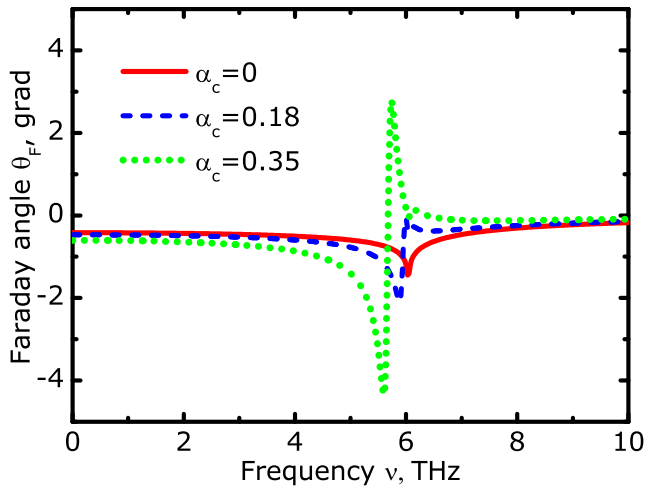}
\caption{(Color online) Frequency dependence of Faraday angle $\theta_\mathrm{F}$ for $\alpha_\mathrm{c}=0$ (solid red line), $\alpha_\mathrm{c}=0.18$ (dashed blue line) and $\alpha_\mathrm{c}=0.35$ (dotted green line).}
\end{figure}

If incident electromagnetic wave is linearly polarized $\vec{E}=\vec{e}_x E_0$, where $E_0$ is its amplitude, characteristics of transmitted and reflected waves can be expressed in terms of amplitudes of transmission $t_\lambda=|t_\lambda|e^{i\Phi_\lambda^\mathrm{t}}$ and reflection $r_\lambda=|r_\lambda|e^{i\Phi_\lambda^\mathrm{r}}$ of circularly polarized electromagnetic wave  $\vec{E}=(\vec{e}_x+i \lambda \vec{e}_y) E_0$, where $\lambda=\pm1$ is the sign of the circular polarization. Transmittance of electromagnetic wave through TI film is given by $T=(|t_+|^2+|t_-|^2)/2$. Angle of polarization plane rotation (Faraday angle) $\theta_\mathrm{F}$ and ellipticity $\delta_\mathrm{F}$ of transmitted wave are given by $\theta_\mathrm{F}=(\Phi_+^\mathrm{t}-\Phi_-^\mathrm{t})/2$ and $\delta_\mathrm{F}=(|t_+|-|t_-|)/(|t_+|+|t_-|)$,  respectively. Angle of polarization plane rotation (Kerr angle) $\theta_\mathrm{K}$ and ellipticity $\delta_\mathrm{K}$ of reflected wave are given by $\theta_\mathrm{K}=(\Phi_+^\mathrm{r}-\Phi_-^\mathrm{r})/2$ and $\delta_K=(|r_+|-|r_-|)/(|r_+|+|r_-|)$, respectively.

Using Maxwell equations for thin film geometry and the boundary conditions for electric and magnetic fields that take into account electric currents excited by electromagnetic waves we find
\begin{equation}
t_\lambda=\frac{\sigma_0}{\sigma_0+\alpha \sigma_\lambda^\mathrm{t}}, \quad \quad r_\lambda=-\frac{\alpha \sigma_\lambda^\mathrm{t}}{\sigma_0+\alpha \sigma_\lambda^\mathrm{t}},
\end{equation}
where $\sigma_0=e^2/h$ is quantum of conductivity; $\alpha\approx1/137$ is fine structure constant; $\sigma_\lambda^\mathrm{t}=\sigma_{xx}^\mathrm{t}+i\lambda\sigma_{yx}^\mathrm{t}$; $\sigma_{\alpha\beta}^\mathrm{t}$ is sum of the optical conductivity components from the opposite surfaces of the TI film. Direction of polarization plane rotations for reflected and transmitted electromagnetic waves depends on sign the of the Hall conductivity and, hence, on sign of $z$-component of exchange field. The rotations on the opposite surfaces of the TI film enhance each other if the sign of the corresponding component of exchange fields on opposite surfaces coincide. The case is considered below. If also the absolute values of the magnetically induced gaps on the opposite surfaces are the same then $\sigma_{\alpha\beta}^\mathrm{t}=2 \sigma_{\alpha\beta}$. If the corresponding components of exchange fields have different signs and the same absolute values than $\sigma_{\alpha\beta}^\mathrm{t}=0$ and  Faraday and Kerr effects vanish.

\begin{figure}[t]
\label{Fig6}
\includegraphics[width=8.5cm]{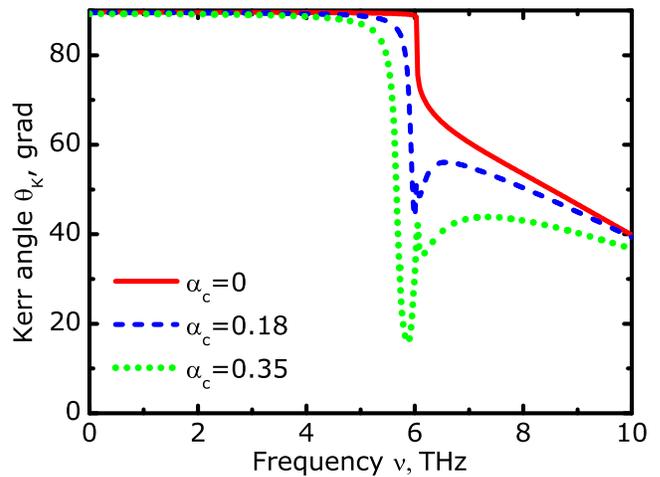}
\caption{(Color online) Frequency dependence of Kerr angle $\theta_\mathrm{K}$ for $\alpha_c=0$ (solid red line), $\alpha_c=0.18$ (dashed blue line) and $\alpha_c=0.35$ (dotted green line).}
\end{figure}

Dependence of transmittance $T$ of electromagnetic wave through TI thin film on its frequency  $\omega$ is represented in Fig. 4. Chiral excitonic levels lead to resonant absorption of energy of electromagnetic wave. But absorption of energy is too small to prevent detection of transmitted electromagnetic wave\cite{Comment1}.

Dependence of Faraday angle $\theta_\mathrm{F}$ and Kerr angle $\theta_\mathrm{K}$  on the frequency $\omega$ is presented in Fig. 5 and Fig. 6. At low frequencies contribution of excitons to optical conductivity is insignificant and Faraday $\theta_\mathrm{F}$ and Kerr angles $\theta_\mathrm{K}$ tend to their universal values  $\tan \theta_\mathrm{F}=\alpha$ and $\tan \theta_\mathrm{K}=1/\alpha$, respectively, where $\alpha\approx1/137$ is fine structure constant. Resonance enhancement of Hall conductivity by chiral excitons leads to resonant enhancement of Faraday angle.  Kerr angle is very sensitive to longitudinal component of the conductivity tensor\cite{TkachovHankiewicz}. At the resonant condition longitudinal component of the optical conductivity has sharp peak and, hence, Kerr angle is considerably reduced. Both of these prominent effects can be directly observed in experiment.

Dependence of the ellipticities of transmitted $\delta_\mathrm{F}$ and reflected $\delta_\mathrm{K}$ electromagnetic waves are presented in Fig. 7 and Fig. 8. Chiral excitons resonantly enhance ellipticity of transmitted wave $\delta_\mathrm{F}$ and lead to observable signature in ellipticity $\delta_\mathrm{K}$ of reflected wave. These signatures can also be directly observed in the experiments.

\section{Conclusions}
Resonant features in  Faraday and Kerr effects caused by chiral excitons in thin TI film can be observable in experiments if contribution of at least single excitonic level to Hall conductivity is well separated from single-particle contribution. Thus the excitonic binding energy $\Omega_{\mathrm{b}}=\alpha_\mathrm{c}^2 |\Delta|$ of the lowest energy state $|0,1\rangle$ should exceed the excitonic decay rate $\gamma$. The maximal value of the gap induced in $\hbox{Bi}_2 \hbox{Se}_3$ by ordered magnetic impurities\cite{TIExchange1} is $2|\Delta|\approx50\; \hbox{meV}$. For $\hbox{Bi}_2 \hbox{Se}_3$ dimensionless coupling constant and the binding energy are $\alpha_\mathrm{c}=0.09$ and $\Omega_{\mathrm{b}}\sim 2 \hbox{K}$. For $\hbox{Bi}_2 \hbox{Te}_3$ dimensionless coupling constant and the binding energy are $\alpha_\mathrm{c}=0.18$ and $\Omega_{\mathrm{cr}}\sim 9 \hbox{K}$. Exciton decay rate $\gamma$ can be estimated from scattering rate of electrons. Maximal value of the electron mobility in absence of the gap estimated from transport experiments\cite{QiZhang} is $\mu \sim 10^4\; \hbox{cm}^2/\hbox{eV}\cdot \hbox{s}$. It corresponds to scattering rate $\gamma\sim 30 \hbox{K}$, hence, present parameters relevant to the experiments are close enough to the favorable ones.

\begin{figure}[t]
\label{Fig7}
\includegraphics[width=8.5cm]{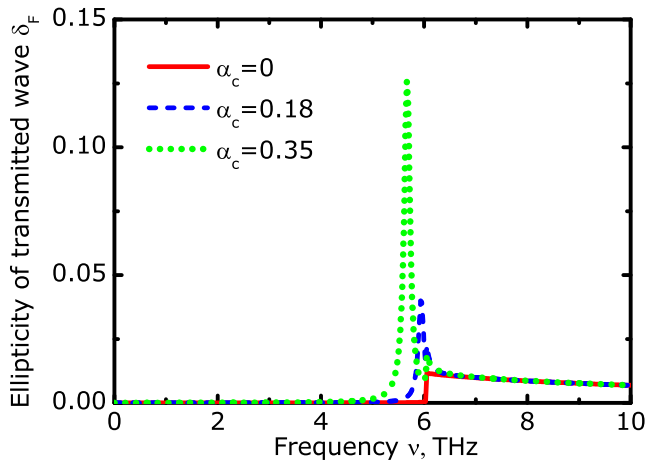}
\caption{(Color online) Frequency dependence of ellipticity of transmitted wave $\delta_\mathrm{F}$ for $\alpha_\mathrm{c}=0$ (solid red line), $\alpha_\mathrm{c}=0.18$ (dashed blue line) and $\alpha_\mathrm{c}=0.35$ (dotted green line).}
\end{figure}

The dielectric constant of bismuth and telluride based TI achieves $\varepsilon\approx 40-100$ hence coupling constant and exciton binding energies of the chiral exciton energies are rather small. So resonant excitonic resonances are fragile to disorder and finite temperature effects. The problem can be overcome in ultrathin TI films which width is considerably less then exciton radius $k_{\mathrm{exc}}d\ll1$. In that case the effective dielectric constant of a TI film equals to half-sum of dielectric constant of a substrates surrounding the film and does not depend on one of TI. Its value can be considerably smaller than dielectric permittivity of TI. For  $\alpha_\mathrm{c}=0.18$ character value of $d$ equals to $60\; \hbox{nm}$.

We used Hartree-Fock approximation for the description of the delocalized single-particle states. In this approximation correlation between motion of electron and hole is neglected and their wave functions are approximated by independent plane waves. So Coulomb interaction leads only to renormalization of single-particle spectrum. Correlations for delocalized electron-hole states (unbound excitons)  in more complicated approximation (that involves the calculation of two-particle Green function) were considered for conventional semiconductor nanostructures (see \cite{Elliot,Glutch} and references therein). Coulomb interaction leads to enhancement of absorption spectrum with Sommerfeld-Gamov factor. But in any case single-particle contribution is not resonant one and it is smooth as function of frequency. Hence if the chiral excitonic state on a TI surface is well separated from continuum of single-particle transitions ($\omega\ge2|\Delta|$) the resonant  Faraday and Kerr effects will be observable.

\begin{figure}[t]
\label{Fig8}
\includegraphics[width=8.5cm]{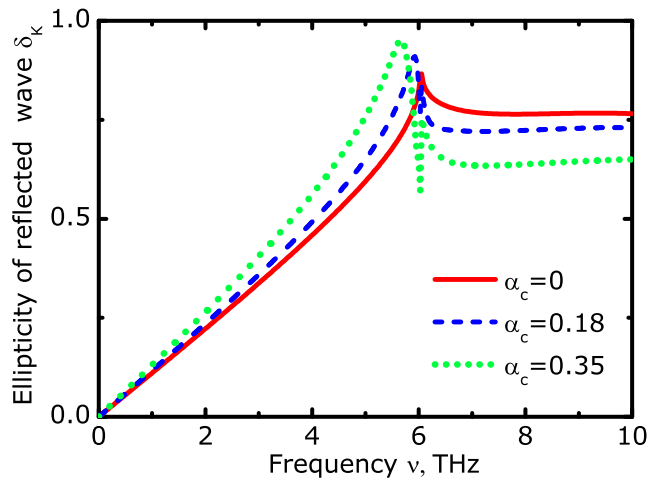}
\caption{(Color online) Frequency dependence of ellipticity of reflected wave $\delta_\mathrm{K}$ for $\alpha_\mathrm{c}=0$ (solid red line), $\alpha_\mathrm{c}=0.18$ (dashed blue line) and $\alpha_\mathrm{c}=0.35$ (dotted green line).}
\end{figure}

Chiral excitons also appear\cite{ParkLouie} in monolayer and bilayer graphene. In the former the gap in the spectrum can be induced by the special substrate, for example $\hbox{BN}$\cite{GrapheneGapBN} or $\hbox{SiC}$\cite{GrapheneGapSiC}. The spectrum in latter can be gapped by external electric field perpendicular to layer\cite{BilayerGapTheor,BilayerGapExp1,BilayerGapExp2}. The chiral excitonic state with minimal energy in monolayer and bilayer graphene has orbital angular momentum $m=1$ and $m=2$, respectively. In the first Brillouin zone of both materials there are two valleys connected with each other by the time reversal transformation \cite{GrapheneReview}. Due to the time reversal symmetry the value of the gaps in two valleys have exactly the same absolute values and opposite signs $|\Delta|$ and $-|\Delta|$. The chiral excitons from single valley resonantly contribute to the optical Hall conductivity but the contributions of two valleys cancel each other. Therefore the chiral excitons in gapped monolayer and bilayer graphene do not manifest in Faraday and Kerr effect in absence of external magnetic field.

The time reversal symmetry on a surface of TI can be also broken by external magnetic field perpendicular to the surface leading to reconstruction of the Dirac spectrum to separate Dirac Landau levels\cite{HasanKane,QiZhang}. Optical Hall conductivity consists of the set of resonances that correspond to optical single-particle transitions between Dirac Landau levels\cite{TseMacDonaldMF}. In presence of Coulomb interaction energy of single-particle transition depends on its total momentum and the set of Landau levels transforms to the set of the dispersive magnetoexcitonic branches \cite{KallinHalperin,LernerLozovik1,LernerLozovik2,DzubenkoLozovik,Moscalenko,Koinov,PikalovFil,LozovikSokolik}. Only magnetoexcitons with zero total momentum contribute to the optical conductivity tensor hence Coulomb interaction does not qualitatively change frequency dependence of Hall conductivity.  Coulomb interaction shifts the positions of the resonances and can change their amplitudes. Manifestation of magnetoexcitons in  Faraday and Kerr effects is not so prominent as chiral excitons on a TI surface with the magnetically induced gap.

At present Faraday and Kerr effects in thin TI films subjected to external perpendicular magnetic field are extensively studied experimentally \cite{JenkinsSushkovSchmadel,SobotaYangAnalytis,HancockMechelenKuzmenko,AguilaStierLiu}.
Real samples contain residual bulk charge carriers (that are not completely excluded by doping) and polar phonons interacting with electromagnetic waves. Complicated dependencies of Kerr and Faraday angles and peaks in longitudinal conductivity observed in \cite{JenkinsSushkovSchmadel,SobotaYangAnalytis,HancockMechelenKuzmenko,AguilaStierLiu} are interpreted in terms of bulk response. Observed signatures that do not depend on TI film thickness can be caused by
magnetoexcitons.

 In summary, we theoretically investigated the manifestations of chiral excitons on the magnetically gapped surfaces of topological insulator film in  Faraday and Kerr effects. Contribution of chiral excitons to optical conductivity tensor of the TI surface is calculated. As conventional excitons chiral one lead to sharp peak of longitudinal conductivity and to resonance absorption of energy of incident electromagnetic wave. Contrary to conventional excitons chiral ones due to lack of symmetry between states with opposite angular momentum resonantly enhance Hall conductivity and play important role in  effect. Chiral excitons lead to considerable enhancement of Faraday angle and ellipticity of transmitted electromagnetic wave at resonance condition. Also they lead to resonant weakening of Kerr angle and prominent signatures in ellipticity of reflected electromagnetic wave. The described effects can be directly observed in the experiments.

\begin{acknowledgments}

The authors are indebted to A.A. Sokolik for fruitful discussions. The work was supported by RFBR programs. D.K.E acknowledge support from Grant of the President of Russian Federation MK-5288.2011.2, RFBR grant 12-02-31199 and by scholarship from Dynasty Foundation.

\end{acknowledgments}

\bibliographystyle{apsrev}

\begin{thebibliography}{99}
\bibitem{Comment1}
In real samples $\hbox{Bi}_2 \hbox{Se}_3$ and $\hbox{Bi}_2 \hbox{Te}_3$ there are inevitable bulk charge carrires and polar phonons that also lead to absorbtion of energy.
\end{thebibliography}

\end{document}